\documentclass[a4paper,aps, twocolumn,prl,amsmath,amssymb,superscriptaddress, 12p6]{revtex4-1}

\usepackage{graphicx}
\usepackage{dcolumn}
\usepackage{bm}
\usepackage{longtable}
\usepackage{color}
\usepackage{ulem}

\makeatletter
\newcommand\figcaption{\def\@captype{figure}\caption}
\newcommand\tabcaption{\def\@captype{table}\caption}

\makeatother

\begin{document}

\title{Microscopic origin of the mobility enhancement at a spinel/perovskite oxide heterointerface revealed by photoemission spectroscopy}

\author{P. Sch{\"u}tz}
\affiliation{Physikalisches Institut and R\"ontgen Center for Complex Material Systems (RCCM), Universit\"at W\"urzburg, Am Hubland,
D-97074 W\"urzburg, Germany}
\author{D.V. Christensen}
\affiliation{Department of Energy Conversion and Storage, Technical University of Denmark, 4000 Roskilde, Denmark}
\author{V. Borisov}
\affiliation{ Institute of Theoretical Physics, Goethe University Frankfurt am Main, D-60438 Frankfurt am Main, Germany}
\author{F. Pfaff}
\author{P. Scheiderer}
\author{L. Dudy}
\author{M. Zapf}
\author{J. Gabel}
\affiliation{Physikalisches Institut and R\"ontgen Center for Complex Material Systems (RCCM), Universit\"at W\"urzburg, Am Hubland,
D-97074 W\"urzburg, Germany}
\author{Y.Z. Chen}
\author{N. Pryds}
\affiliation{Department of Energy Conversion and Storage, Technical University of Denmark, 4000 Roskilde, Denmark}
\author{V.A. Rogalev}
\affiliation{Physikalisches Institut and R\"ontgen Center for Complex Material Systems (RCCM), Universit\"at W\"urzburg, Am Hubland,
D-97074 W\"urzburg, Germany}
\affiliation{Swiss Light Source, Paul Scherrer Institut, CH-5232 Villigen, 
Switzerland}
\author{V.N. Strocov}
\affiliation{Swiss Light Source, Paul Scherrer Institut, CH-5232 Villigen, Switzerland}
\author{C. Schlueter}
\author{T.-L. Lee}
\affiliation{Diamond Light Source, Harwell Sciene and Innovation Campus, Oxfordshire OX11 0DE, United Kingdom}
\author{H.O. Jeschke}
\author{R. Valent{\'i}}
\affiliation{ Institute of Theoretical Physics, Goethe University Frankfurt am Main, D-60438 Frankfurt am Main, Germany}
\author{M. Sing}
\author{R. Claessen}
\affiliation{Physikalisches Institut and R\"ontgen Center for Complex Material Systems (RCCM), Universit\"at W\"urzburg, Am Hubland,
D-97074 W\"urzburg, Germany}

\date{\today}
\begin{abstract}%
The spinel/perovskite heterointerface $\gamma$-Al$_2$O$_3$/SrTiO$_3$ hosts a two-dimensional electron system (2DES) with electron mobilities exceeding those in its all-perovskite counterpart LaAlO$_3$/SrTiO$_3$ by more than an order of magnitude despite the abundance of oxygen vacancies which act as electron donors as well as scattering sites. By means of resonant soft x-ray photoemission spectroscopy and \textit{ab initio} calculations we reveal the presence of a sharply localized type of oxygen vacancies at the very interface due to the local breaking of the perovskite symmetry. We explain the extraordinarily high mobilities by reduced scattering resulting from the preferential formation of interfacial oxygen vacancies and spatial separation of the resulting 2DES in deeper SrTiO$_3$ layers. Our findings comply with transport studies and pave the way towards defect engineering at interfaces of oxides with different crystal structures.
\end{abstract}
\maketitle
\begin{figure*}%
\centering
\includegraphics[width = 0.9\linewidth]{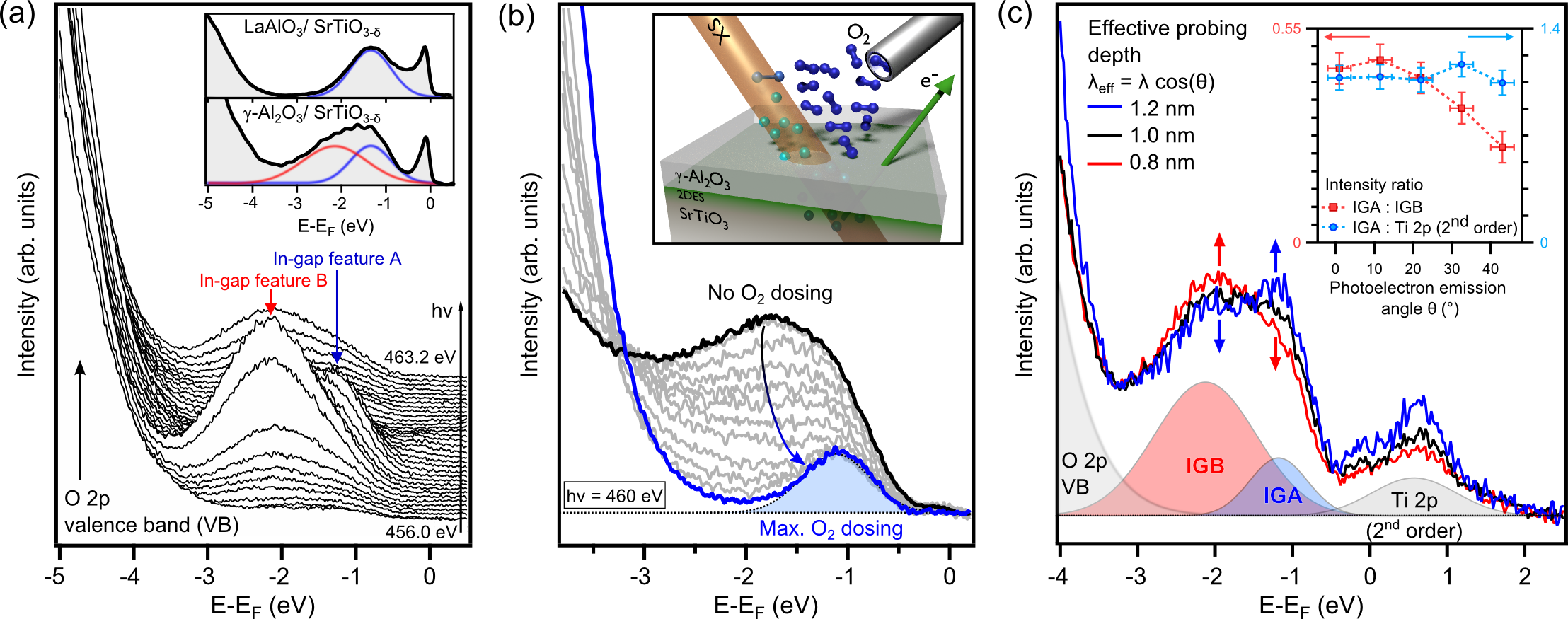}%
\caption{(a) Photoemission spectra of a $\gamma$-Al$_2$O$_3$/SrTiO$_3$ heterostructure upon tuning the photon energy across the Ti~$L$ absorption edge. Two in-gap features, IGA and IGB, are distinguished by their shifted resonance photon energies. Inset: Comparison between the systems LaAlO$_3$/SrTiO$_{3-\delta}$ and $\gamma$-Al$_2$O$_3$/SrTiO$_{3-\delta}$ ($h\nu = 459.4\,\text{eV}$). (b) Resonant Ti~$3d$ spectral weight upon oxygen dosing (schematically depicted in the inset). While IGB becomes quenched entirely, finite IGA weight remains at maximum dosing. (c) Depth-resolved photoemission spectroscopy of the Ti~$3d$ in-gap states and the 2$^\text{nd}$ order light-induced Ti~$2p$ core level ($h\nu = 459.8\,\text{eV}$). The spectra are normalized to the spectral weight between $E=-0.5\,\text{eV}$ and $-3.5\,\text{eV}$. Inset: Relative spectral weight as function of photoemission angle. The strong IGA-to-IGB angle dependence evidences the sharp vertical separation of IGA and IGB.}%
\label{fig:Grafik2}
\end{figure*}

The search for high-mobility two-dimensional electron systems  (2DES) at atomically engineered transition metal oxide heterointerfaces is an ongoing endeavor, since the strong electronic correlations in partially occupied $d$-orbitals promise an even richer physical behavior than found in conventional semiconductor heterostructures \cite{mannhart_oxide_2010}. However, the charge carrier mobilities in the most prominent complex oxide 2DES, found at the perovskite-perovskite heterointerface between the band insulators LaAlO$_3$ and SrTiO$_3$, still fall short of those in semiconductors by several orders of magnitude \cite{thiel_tunable_2006}. The hitherto-highest mobility in SrTiO$_3$-based structures (140,000\,$\text{cm}^2\text{/Vs}$ at $2\,\text{K}$) is found at the spinel/perovskite heterointerface between $\gamma$-Al$_2$O$_3$ thin films and SrTiO$_3$ \cite{chen_high-mobility_2013,chen_room_2013}, thus making it a promising candidate for applications in oxide electronics or fundamental studies of quantum transport.\\
The mechanism of 2DES formation in SrTiO$_3$-based heterostructures has been under debate for many years. The existence of a critical film thickness for metallicity at the epitaxial LaAlO$_3$/SrTiO$_3$ heterointerface has been associated with the polar discontinuity at the interface and the concomitant build-up of an electrostatic potential gradient as a function of film thickness, which may be countered by a charge redistribution to the interface \cite{nakagawa_why_2006, chen_high-mobility_2013, schutz_band_2015, yu_polarity-induced_2014, scheiderer_surface-interface_2015}. Additionally, substantial oxygen vacancy (V$_\text{O}$) doping within the SrTiO$_3$ substrate may occur, which is the dominant source of charge carriers in 2DESs discovered in SrTiO$_3$ interfaced with amorphous overlayers \cite{ herranz_high_2012,  liu_origin_2013} and bare SrTiO$_3$ surfaces irradiated with ultraviolet light \cite{santander-syro_two-dimensional_2011, meevasana_creation_2011, walker_carrier-density_2015}. In the case of $\gamma$-Al$_2$O$_3$/SrTiO$_3$, it has been argued that its 2DES originates exclusively from oxygen vacancies in SrTiO$_3$, which are formed due to redox reactions \cite{chen_high-mobility_2013}. However, it remains unclear how high mobilities can be achieved, when SrTiO$_3$ hosts both conduction electrons and oxygen vacancies, which act as scattering sites. Here, we reveal the existence of a specific type of oxygen vacancies at the spinel/perovskite $\gamma$-Al$_2$O$_3$/SrTiO$_3$ interface by means of soft x-ray resonant photoemission spectroscopy (SX-ResPES) and \textit{ab initio} calculations and propose a spatial separation of the dopants (oxygen vacancies) and the 2DES as the origin for its exceptionally high mobility.\\%
\begin{figure*}%
\centering%
\includegraphics[width =  \linewidth]{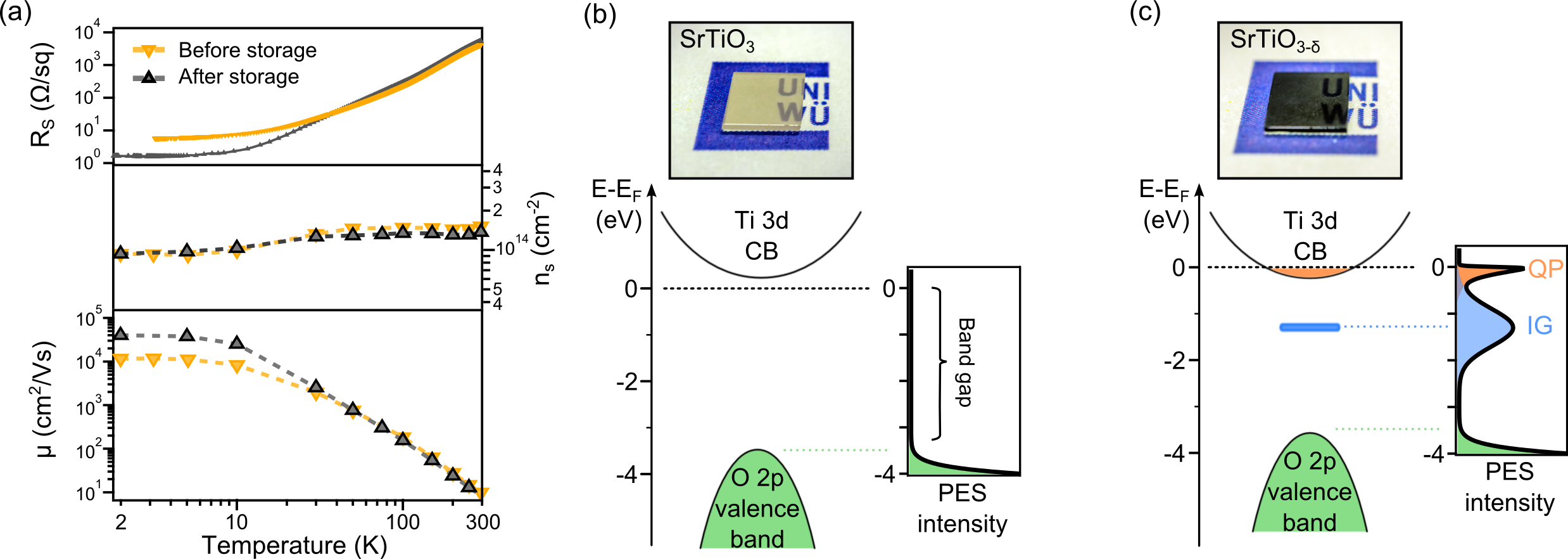}%
\caption{(a) Sheet resistivity ($R_S$), carrier density ($n_S$) and mobility ($\mu$) as function of temperature $T$ of a $\gamma$-Al$_2$O$_3$/SrTiO$_3$ heterostructure before and after 6 months of storage at room temperature in a vacuum desiccator. (b) Photograph, electronic structure and corresponding schematic photoemission spectrum of stoichiometric SrTiO$_3$ and (c) oxygen-deficient SrTiO$_{3-\delta}$.} %
\label{fig:Grafik1}%
\end{figure*}%
SX-ResPES experiments were performed at the combined soft and hard x-ray beamline I09, Diamond Light Source and at the soft x-ray beamline ADRESS, Swiss Light Source \cite{strocov_high-resolution_2010, strocov_soft-x-ray_2013}. Theoretical calculations were performed based on the density functional theory in the generalized gradient approximation (GGA) \cite{perdew_generalized_1996, blochl_projector_1994, kresse_ultrasoft_1999, kresse_efficient_1996,hafner_ab-initio_2008, liechtenstein_density-functional_1995, koepernik_full-potential_1999, okamoto_lattice_2006,bellaiche_virtual_2000} by explicitly including oxygen vacancies in a $\gamma$-Al$_2$O$_3$/SrTiO$_3$ superlattice and a SrTiO$_3$ supercell. For theoretical and experimental details including sample fabrication see Supplemental Material \cite{Supplemental}.
While the $\gamma$-Al$_2$O$_3$/SrTiO$_3$ heterointerface can host a 2DES with extraordinarily high electron mobilities upon careful control of deposition parameters \cite{chen_high-mobility_2013}, for a wide range of growth conditions a lower-mobility 2DES forms \cite{chen_room_2013}. We have found that such lower mobilities can be increased post-growth by gentle low-temperature annealing or prolonged storage at room temperature as shown in Fig.~\ref{fig:Grafik1}(a). The electron density of a $\gamma$-Al$_2$O$_3$/SrTiO$_3$ sample obtained from standard Hall measurements in the van der Pauw geometry \cite{chen_high-mobility_2013} remains largely unaffected in this process. In contrast, the mobility at $2\,\text{K}$ and the residual resistance ratio ($R_S(300\,\text{K})/R_S(2\,\text{K})$) increase by roughly a factor of 4 ($12,000$ to $40,000\,\text{cm}^2\text{/Vs}$ and $730$ to $3200$, respectively). \\%
Since the movement of larger cations is frozen at room temperature \cite{gunkel_transport_2014}, cation diffusion can be excluded as driving mechanism. In contrast, diffusion of lighter oxygen anions (and hence oxygen vacancies) is feasible and may affect the transport properties at cryogenic temperatures, where electron scattering by ionized donors is the dominant effect \cite{verma_intrinsic_2014}. Since the total amount of oxygen vacancies, as reflected in the charge carrier density, remains constant, one may speculate that a gradual redistribution of oxygen vacancy scattering sites away from the confined 2DES may be the cause for the mobility enhancement. \\
As shown in Fig.~\ref{fig:Grafik1}(b), nominally stoichiometric SrTiO$_3$ is an intrinsically $n$-doped wide-gap semiconductor with the Fermi energy pinned close to the conduction band minimum \cite{ertekin_interplay_2012}. Upon introduction of oxygen vacancies, two electrons per vacancy become released and the electronic structure of SrTiO$_3$ changes as depicted in Fig.~\ref{fig:Grafik1}(c). A fraction of donor electrons becomes trapped in a localized Ti~$3d$-derived state next to the vacancy site (blue) and the rest is donated into the Ti~$3d$ conduction band (orange) to become itinerant \cite{lin_orbital_2012, lin_electron_2013}. The resulting photoemission spectrum comprises a dispersive feature cut off by the Fermi-Dirac distribution at the chemical potential, the quasiparticle peak (QP), and a broad non-dispersive in-gap peak (IG), which can be seen as hallmark for the presence of oxygen vacancies in SrTiO$_3$ \cite{aiura_photoemission_2002, mckeown_walker_control_2014, berner_direct_2013}.\\
Here, we use synchrotron-based soft x-ray resonant photoemission spectroscopy (SX-ResPES) at the Ti $L$ edge to enhance the spectroscopic contrast for the Ti~$3d$-derived states at the buried interface \cite{drera_spectroscopic_2011,  drera_intrinsic_2014}. The inset of Fig.~\ref{fig:Grafik2}(a) shows the PES spectra of LaAlO$_3$/SrTiO$_{3-\delta}$ and $\gamma$-Al$_2$O$_3$/SrTiO$_{3-\delta}$ at the resonance condition. Despite an overall similarity, i.e., the presence of a QP and an IG peak, the latter is significantly broader and asymmetric for $\gamma$-Al$_2$O$_3$/SrTiO$_{3-\delta}$, signaling a superposition of (at least) two peaks. Indeed, upon tuning the photon energy across the Ti~$L$ absorption edge as shown in Fig.~\ref{fig:Grafik2}(a), two peaks at $E\,\approx\,-1.2\,\text{eV}$ (in-gap feature A, IGA) and $-2.1\,\text{eV}$ (IGB) can be distinguished by their shifted resonances ($h\nu=458.2\,\text{eV}$ and $459.1\,\text{eV}$ for IGB and IGA, respectively). Note that in these spectra the quasiparticle was suppressed by a specific choice of the measurement geometry \cite{Supplemental}.\\
It is well known that the exposure of maximal-valence transition metal oxides like SrTiO$_3$ to high-intensity synchrotron light causes oxygen vacancy formation \cite{knotek_ion_1978}. This can be counteracted by simultaneous dosing with molecular oxygen, which dissociates in the high-intensity x-ray beam and diffuses as atomic species into the substrate where it annihilates vacancies \cite{walker_carrier-density_2015, dudy_situ_2016}.  As shown in Fig.~\ref{fig:Grafik2}(b), using a metal capillary to direct oxygen onto the sample IGB can be quenched entirely, whereas IGA is reduced, yet remains finite even for the highest possible oxygen flow. Thus, we conclude that both IG peaks represent oxygen vacancy sites, but apparently of different kinds. The different susceptibility to oxygen dosing indicates a different vertical spatial distribution of the two kinds of vacancies,  i.e., the completely quenched IGB has its origin in closer proximity to the interface. Note that the shifted O~$2p$ valence band onset is due to variable band bending in SrTiO$_3$ and a concomitant change of the band alignment at the interface.\\%

\begin{figure*}%
\includegraphics[width =  \linewidth]{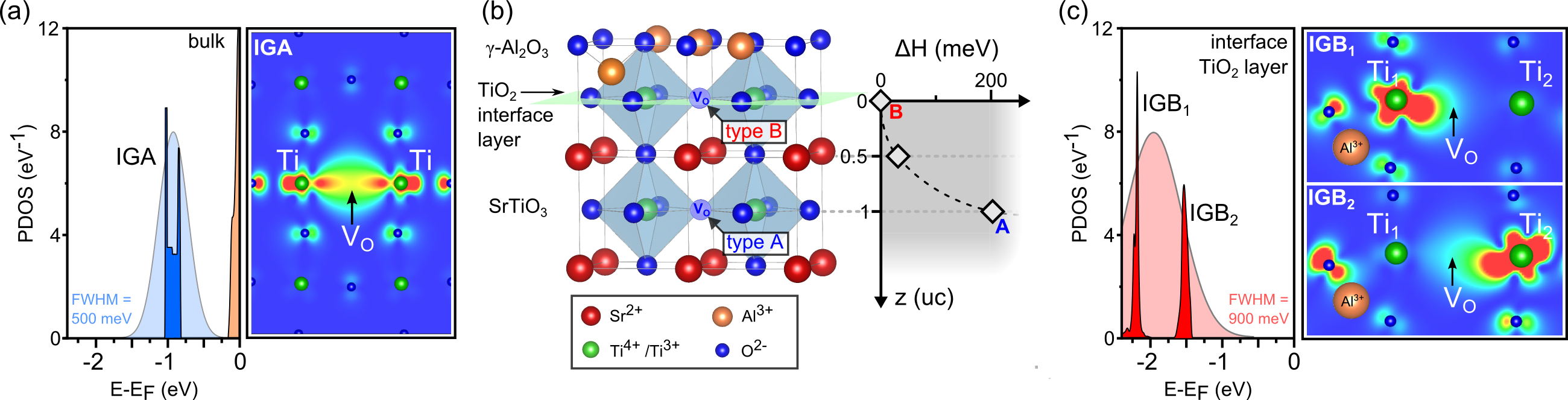}%
\caption{(a) GGA+$U$ calculated Ti~$3d$-projected density of states (PDOS) and real-space electron density of bulk SrTiO$_3$ with an isolated oxygen vacancy. The real-space charge map is shown for a (100) layer cutting the oxygen vacancy and obtained through energy integration of the IGA state. The experimentally observed peak is depicted as guide-to-the-eye curve ($\text{FWHM}=500\,\text{meV}$). (b) Local ionic coordination of bulk-like (type A) and interfacial (type B) oxygen vacancies in an idealized, unrelaxed spinel/perovskite heterointerface and calculated (relative) formation enthalpy $\Delta H$ for oxygen vacancies at and away from the interface. (c) Ti~$3d$-PDOS and real-space electron density of the $\gamma$-Al$_2$O$_3$/SrTiO$_3$ interface with one oxygen vacancy in the TiO$_2$ layer. The real-space charge density is shown for the interfacial TiO$_2$ layer in direct proximity to the spinel film and obtained through energy integration of the resulting IGB$_1$ and IGB$_2$ state, respectively.}%
\label{fig:Grafik3}
\end{figure*}

\begin{figure*}%
\centering
\includegraphics[width =  0.9\linewidth]{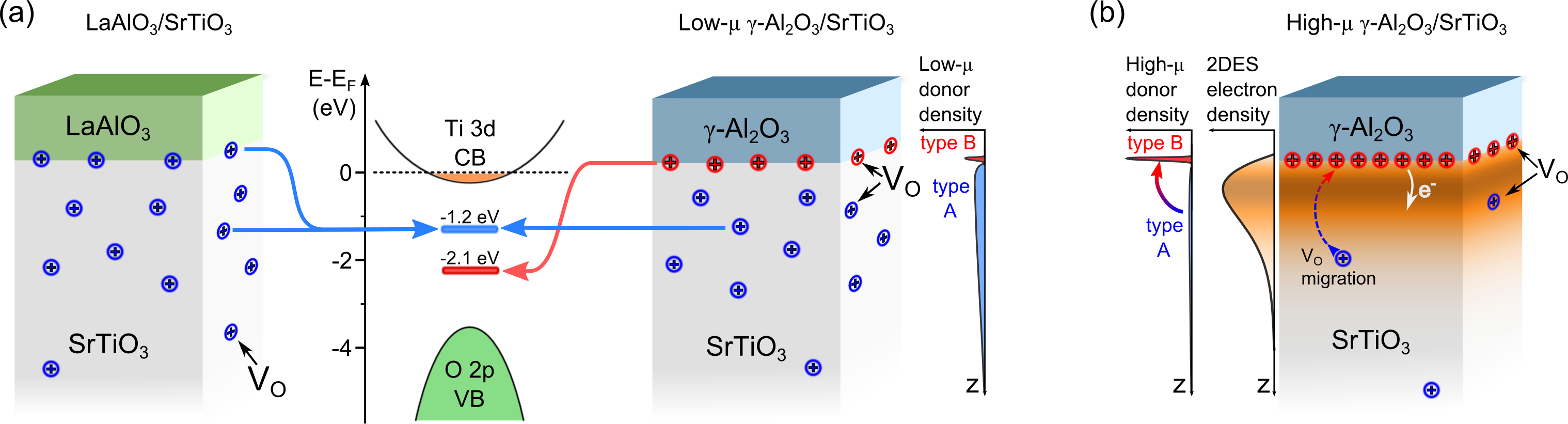}%
\caption{(a) Schematic comparison of oxygen vacancy distribution and energetics in LaAlO$_3$/SrTiO$_3$ and low-mobility $\gamma$-Al$_2$O$_3$/SrTiO$_3$. Possible intrinsic doping mechanisms associated with the polar discontinuity have been omitted here. (b) Proposed mechanism for high mobility conductivity in $\gamma$-Al$_2$O$_3$/SrTiO$_3$ heterostructures. The interfacial oxygen vacancies act as dopant layer, which is spatially separated from the 2DES in the deeper-lying layers of SrTiO$_3$, where ionized donor scattering becomes minimized. Gradual oxygen vacancy diffusion towards the interface during annealing or room-temperature storage changes the oxygen vacancy density from the depicted low-$\mu$ into a high-$\mu$ density distribution.}%
\label{fig:Grafik4}%
\end{figure*}

By acquiring the photoemission signal under various emission angles $\theta$ with respect to the surface normal, the effective probing depth can be changed as $\lambda_{\text{eff}} = \lambda \cos \theta$ \cite{sing_profiling_2009}, where $\lambda$ is the inelastic mean free path. Figure~\ref{fig:Grafik2}(c) shows the resonant Ti~$3d$ spectral weight as a function of $\lambda_\text{eff}$, ranging from $1.2\,\text{nm}$ (maximum bulk sensitivity, $\theta = 0^\circ$) to $0.8\,\text{nm}$ (maximum interface sensitivity, $\theta = 45^\circ$) \cite{tanuma_calculation_2003}. Additionally, the Ti~$2p$ core level signal excited by the second-order light from the undulator is seen above the chemical potential and serves as bulk titanium reference signal. The relative spectral weights were extracted from Gaussian fits and are shown in the inset as a function of emission angle. The considerable angle-dependence of the IGA-to-IGB-intensity ratio (red) provides evidence for an extraordinarily sharp, vertical separation of the two types of oxygen vacancies, i.e., IGB is situated closer to the interface. Furthermore, IGA scales with the second-order light induced Ti~$2p$ bulk signal (blue), indicating a uniform distribution of IGA states throughout the SrTiO$_3$ substrate (within the information depth of roughly $1.2\,\text{nm}$). We thus readily identify IGA as signal from bulk-like oxygen vacancies situated within the substrate and IGB as a fingerprint of oxygen vacancies at the interface. The existence of \textit{two} distinct types of oxygen vacancies--derived here from PES in a dynamical equilibrium situation inherently different from the near-thermodynamical state probed in transport experiments--is nonetheless of high relevance also for the latter situation, as we sketch out in the following.\\
In a qualitative model, the introduction of a single oxygen vacancy into bulk SrTiO$_3$ results in one itinerant electron in  the Ti~$3d\,t_{2g}$ band and one trapped electron in a bonding state derived from adjacent Ti~$3d\,e_g$/$4p_z$ hybrid states \cite{lin_orbital_2012, lin_electron_2013, jeschke_localized_2015}. Indeed, as shown in Fig.~\ref{fig:Grafik3}(a), the Ti~$3d$-projected density of states (PDOS) obtained from GGA+$U$ calculations of a $3\times3\times3$ SrTiO$_3$ supercell with one oxygen vacancy, using standard parameters $U=5\,\text{eV}$ and $J_\mathrm{H}=0.64\,\text{eV}$, exhibits metallic carriers and an IG state at -1\,eV, in good agreement with experiment and the qualitative model. The real-space charge density for a (100) layer cutting through the vacancy and obtained from energy integration of the in-gap state shows a typical Ti~$3d\,e_g$-based bonding orbital \cite{lin_orbital_2012}, which is strongly localized at the oxygen vacancy and its adjacent Ti atoms.\\%
In SrTiO$_3$-based all-perovskite heterostructures like LaAlO$_3$/SrTiO$_3$ all oxygen vacancies in the substrate can be regarded as bulk-like to a good approximation. In contrast, the perovskite symmetry is broken at the $\gamma$-Al$_2$O$_3$/SrTiO$_3$ heterointerface since $\gamma$-Al$_2$O$_3$ adopts a spinel crystal structure \cite{zhou_structures_1991, ealet_electronic_1994, mo_electronic_1997}. As seen in Fig.~\ref{fig:Grafik3}(b), here, the local ionic coordination of an oxygen vacancy in the uppermost TiO$_2$ layer (type B) differs significantly from that in deeper, bulk-like layers (type A). The peculiar spinel cation sublattice lifts the local $C_{4\nu}$ symmetry relevant for oxygen vacancies in bulk SrTiO$_3$, likely resulting in states of different orbital composition \cite{cao_anomalous_2016} and binding energy, as supported by a crystal-field analysis \cite{Supplemental}.\\%
Figure~\ref{fig:Grafik3}(c) shows GGA+$U$ calculations of a $\gamma$-Al$_2$O$_3$/SrTiO$_3$ superlattice with one oxygen vacancy in the interfacial TiO$_2$ layer. The Ti~$3d$-projected density of states exhibits two IG peaks (denoted as IGB$_1$ and IGB$_2$) at $E\,\approx\,-2.3~\text{eV}$ and $-1.5~\text{eV}$, that are hosted by the two adjacent inequivalent Ti cations. As expected the real-space charge density maps corresponding to IGB$_1$ and IGB$_2$ show a rather complex $d$ orbital composition that is, however, reminiscent of $e_g$-like orbitals. The experimentally observed broad peak at $\approx\,-2.1~\text{eV}$ (indicated by a guide-to-the-eye curve) results from oxygen vacancy clustering that will occur in any realistic system and leads to a statistical distribution of in-gap states with slightly different binding energies. Note that the formation enthalpy $\Delta H$ of single oxygen vacancies located in different SrTiO$_3$ layers exhibits a minimum at the interface [Fig.~\ref{fig:Grafik3}(b)], hence suggesting a favored formation of type B oxygen vacancies and/or diffusion of vacancies from the bulk to the interface. A recent annealing study of $\gamma$-Al$_2$O$_3$/SrTiO$_3$ indeed identified such diffusion processes, in support of the mechanism proposed here \cite{christensen_controlling_2017}.\\
Figure~\ref{fig:Grafik4}(a) schematically summarizes our experimental and theoretical findings. The peculiar local symmetry-breaking at the spinel/perovskite heterointerface $\gamma$-Al$_2$O$_3$/SrTiO$_3$ results in a unique type of (interfacial) oxygen vacancy neither found in bulk SrTiO$_3$ nor at its perovskite/perovskite counterpart LaAlO$_3$/SrTiO$_3$. In SrTiO$_3$-based heterostructures oxygen vacancies represent an extrinsic source of electrons \cite{huijben_structureproperty_2009, gabel_disentangling_2017}, but also act as strong scatterers for charge carriers when present near the 2DES. As already pointed out by Huijben et al. their deleterious effect on the mobility requires design strategies to remove them from the transport region, e.g., by a post-growth exposure to an oxygen-rich environment at elevated temperature and possibly an incorporation of a SrCuO$_2$ nano-layer that facilitates oxygen surface exchange ($\mu \approx 50,000\,\text{cm}^2\text{/Vs}$) \cite{huijben_defect_2013}. However, in contrast to LaAlO$_3$/SrTiO$_3$, the $\gamma$-Al$_2$O$_3$/SrTiO$_3$ heterointerface becomes insulating when exposed to oxygen at high temperatures \cite{chen_high-mobility_2013}, since its interfacial 2DES essentially stems from oxygen vacancies and lacks an intrinsic component \cite{gunkel_thermodynamic_2016} despite a possible polar discontinuity \cite{schutz_band_2015, christensen_is_2017}. Therefore a different strategy is needed to achieve high mobilities in $\gamma$-Al$_2$O$_3$/SrTiO$_3$.\\%
We argue that as depicted in Fig.~\ref{fig:Grafik4}(a) the oxygen vacancy concentration and distribution in low-mobility $\gamma$-Al$_2$O$_3$/SrTiO$_3$ heterostructures is similar to that in standard LaAlO$_3$/SrTiO$_3$ samples with comparable mobilities ($\mu\approx 1,000\,\text{cm}^2\text{/Vs}$) \cite{huijben_defect_2013,chen_high-mobility_2013}. In contrast, as shown in Fig.~\ref{fig:Grafik4}(b), we propose a different oxygen vacancy distribution in high-mobility $\gamma$-Al$_2$O$_3$/SrTiO$_3$ heterostructures, where the majority of oxygen vacancies resides at the very interface (type B) and effectively acts as a single layer of electron donors. They provide the itinerant electrons that form the spatially much more extended 2DES. Experimental estimates for its depth from the interface range from 15 \cite{schutz_band_2015} to 75~\AA~\cite{yazdi-rizi_infrared_2016}, where the amount of bulk-like (type A) oxygen defects is strongly suppressed. Reminiscent of modulation-doped semiconductor structures, this spatial separation of ionized donor scattering sites and the 2DES results in a significant mobility enhancement and hence reconciles the coexistence of high mobilities with the abundance of oxygen vacancies. Note that an additional contribution to the mobility enhancement may be the increased electronic screening of electron-phonon interactions, that suppresses the formation of polarons with an enhanced effective mass in SrTiO$_3$-based structures \cite{wang_tailoring_2016, chang_structure_2010, cancellieri_polaronic_2016}.\\%
Our model also offers a natural explanation for the extremely narrow parameter window for the formation of a high-mobility 2DES \cite{chen_room_2013}. Pulsed laser deposition (PLD) of $\gamma$-Al$_2$O$_3$ thin films on SrTiO$_3$ results in an oxygen vacancy distribution, which is determined by the intricate interplay between bulk and interface oxygen vacancy formation during the deposition process and by their redistribution during cool-down. Naturally, small deviations from the optimum growth conditions (e.g. oxygen pressure, temperature and laser fluency) will result in oxygen vacancies in the 2DES region, enhanced ionized donor scattering and a lower mobility. Likewise, the post-growth mobility enhancement during room-temperature storage can be explained by gradual oxygen vacancy migration towards the energetically-favorable interface, as depicted in Fig.~\ref{fig:Grafik4}(b), resulting in reduced scattering of the 2DES electrons while conserving the total amount of donors.\\
Oxygen vacancies at the LaAlO$_3$/SrTiO$_3$ heterointerface have been associated with ferromagnetism, i.e., ordered local magnetic moments trapped at neighboring Ti sites \cite{kalisky_critical_2012,pavlenko_magnetic_2012}. We speculate that oxygen vacancies at the spinel-perovskite $\gamma$-Al$_2$O$_3$/SrTiO$_3$ interface may also favor magnetic ordering that could qualitatively differ from their deeper-lying counterparts and LaAlO$_3$/SrTiO$_3$. A careful experimental investigation of the magnetic properties offers a promising path to a more complete insight into the physics of this intriguing system.

\begin{acknowledgments}
This work was supported by the DFG (SFB 1170, SFB/TRR 49, FOR 1162 and FOR 1346) 
and the BMBF (05K13WW1). We thank Giorgio Sangiovanni for fruitful discussions 
and David McCue for his technical support during the DLS I09 beamtimes.
\end{acknowledgments}


%

\end{document}